\documentclass{article}

\usepackage{graphicx}
\usepackage{psfig}
\usepackage{epsfig}
\usepackage[round]{natbib}

\title{Probabilistic temperature forecasting: a comparison of four spread-regression models}

\author{Stephen Jewson\footnote{\emph{Correspondence address}: RMS, 10 Eastcheap,
London, EC3M 1AJ, UK. Email: \texttt{x@stephenjewson.com}}}

\begin{document}

\maketitle

\begin{abstract}
Spread regression is an extension of linear regression that allows
for the inclusion of a predictor that contains information about
the variance. It can be used to take the information from a
weather forecast ensemble and produce a probabilistic prediction
of future temperatures. There are a number of ways that spread
regression can be formulated in detail. We perform an empirical
comparison of four of the most obvious methods applied to the
calibration of a year of ECMWF temperature forecasts for London
Heathrow.
\end{abstract}

\section{Introduction}

There is considerable demand within industry for probabilistic forecasts of
temperature, particularly from industries that routinely use probabilistic
analysis such as insurance, finance and energy.
However there is considerable disagreement among meteorologists about how
such forecasts should be produced and at present
no adequately calibrated probabilistic forecasts are available commercially.
Those who need to use probabilistic forecasts have to make them themselves.

How, then, should probabilistic forecasts of temperature be made?
A number of very different methods have been suggested in the
literature such as those described in~\citet{mylne02a},
\citet{roulstons03} and~\citet{raftery}. However it seems that all
three of these methods, although complex, suffer from the
shortcoming that they don't calibrate the amplitude of variations
in the ensemble spread but rather leave the amplitude to be
determined as a by-product of the calibration of the mean.

We take a very different, and simpler, approach to the development
of probabilistic forecasts than the authors cited above.
Our approach is based on the following philosophy:

\begin{itemize}

    \item The baseline for comparison for all probabilistic temperature forecasts should be a
    distribution derived very simply by using linear regression around a single forecast or an ensemble mean.

    \item More complex methods can then be tested against this baseline.
    Before anything more complex than linear regression
    is adopted on an operational basis it should be shown
    to clearly beat linear regression in out of sample tests.
    Unfortunately none of the studies cited above compared the methods they proposed
    with linear regression, and, given that they seem not to calibrate the ensemble spread
    correctly, it would seem possible that they might not perform as well.

\end{itemize}

We have followed this philosophy and, based on our analysis of one particular dataset of
past forecasts and past observations we have shown that:

\begin{itemize}

    \item Moving from constant-parameter linear regression to seasonal parameter linear
    regression gives a huge improvement in forecast skill for forecasts of both the mean
    temperature and the distribution of temperatures~\citep{jewson04c}

    \item Adding spread as a predictor gives only a very small improvement
    (\citet{jewsonbz03a}, \citet{jewson03g}).

    \item Generalising to allow for non-normality gives no improvement at all~\citep{jewson03h}.

\end{itemize}

All these results are summarised and discussed in~\citet{jewson04l}.

In this article we focus on the second of these conclusions: that using the spread
as an extra predictor brings only a very small improvement to forecast skill.
This is somewhat disappointing given that it had been hoped by some that use of the
ensemble spread would turn out to be an important factor in the creation of
probabilistic forecasts. We are trying to get a better understanding of \emph{why}
the ensemble spread brings so little benefit in the tests we have performed.
In~\citet{jewson04p} we concluded that this is because of:

\begin{enumerate}

    \item The scoring system we use.\\
    We calibrate and score probabilistic forecasts using the likelihood of
    classical statistics (\citet{fisher1912}, \citet{jewson03d}).
    Likelihood, as we have used it, is a measure that considers the ability of the
    forecast to predict the whole distribution of future temperatures. Much of the mass
    in the distribution of temperature is near the mean and so the likelihood naturally tends
    to emphasize the importance of the mean rather than the spread.
    If we were to use a score that puts more weight onto the tails of the distribution
    then the spread might prove more important (although such a score would not then reflect our main
    interest, which is in the prediction of the whole distribution).

    \item The low values of the coefficient of variation of spread (COVS).\\
    Once we have calibrated our ensemble forecast data we find that the
    uncertainty does not
    vary very much relative to the mean level of the uncertainty (i{.}e{.} the COVS is low).
    Thus if we approximate the uncertainty with a constant this does not degrade the
    forecast to any great extent, and we have not been able to detect a significant
    impact of the spread in out of sample testing.
    That the variations in the calibrated uncertainty are small
    could be either because the actual uncertainty does not vary
    very much or because the ensemble spread is not a good predictor for the actual uncertainty.
    In fact it is likely to be a combination of these two effects.

    \item The low values of the spread mean variability ratio (SMVR).\\
    We have also found that the amplitude of the variations in the uncertainty
    in the calibrated forecast is small relative to the amplitude of the variations in the mean temperature
    (i{.}e{.} the SMVR is low).
    As a result accurate prediction of the (small) variations in the uncertainty is not
    very important relative to accurate prediction of the (large) variations in the mean
    temperature.

\end{enumerate}

However in addition to these reasons
it is also possible that we have been using the ensemble spread wrongly
in our predictions. The model we have been using represents the unknown
uncertainty $\sigma$ as a linear function of the ensemble spread~\citep{jewsonbz03a}:

\begin{eqnarray}
    \sigma&=&\hat{\sigma}+\mbox{noise}\\
          &=&\delta+\gamma s +\mbox{noise}
\end{eqnarray}
But this model is entirely ad-hoc. Why a linear function? We chose
linear because it is the simplest way to calibrate both the
mean uncertainty and the amplitude of the variability of the uncertainty, and not on the
basis of any theory or analysis of the empirical spread-skill relationship.
This suggests it is very important to test other models to see if they perform any better.

In this paper we will compare the original spread-regression model with
3 other spread-regression models.
The four models we compare all have four parameters and
so can be compared in-sample. This is important because the signals
we are looking for are weak and obtaining
long stationary series of past forecasts is more or less impossible at this
point in time.
At some point the numerical modellers will hopefully start providing long
(i{.}e{.} multiyear)
back-test time series from their models. This will allow more
thorough out of sample testing of calibration schemes such as
the spread-regression model
and will facilitate the comparison of models
with different numbers of parameters: meanwhile we do what we can
with the limited data available.

\section{Four spread regression models}

The four spread-regression models that we will test are all based on linear
regression between anomalies of the temperature and anomalies of the ensemble mean:
\begin{equation}
 T_i \sim N(\alpha+\beta m_i,\hat{\sigma})
\end{equation}
The difference between the models is in the representation of $\hat{\sigma}$.

The original standard-deviation-based spread regression model is:
\begin{equation}\label{linearinsd}
 \hat{\sigma}_i=\gamma+\delta s_i
\end{equation}

The variance-based model is:
\begin{equation}
 \hat{\sigma}^2_i=\gamma^2+\delta^2 s_i^2
\end{equation}

The inverse-standard-deviation-based model is:
\begin{equation}
 \frac{1}{\hat{\sigma}_i}=\gamma+\frac{\delta}{s_i}
\end{equation}

and the inverse-variance-based-model is:
\begin{equation}
 \frac{1}{\hat{\sigma}_i^2}=\gamma^2+\frac{\delta^2}{s_i^2}
\end{equation}

Following~\citet{jewson04c} the parameters $\alpha, \beta, \gamma, \delta$
all vary seasonally using a single
sinusoid.
We fit each model by finding the parameters that maximise the likelihood (using
numerical methods).

We note that for very small variations in $s$ all these models can be linearised
and end up the same as the linear-in-standard-deviation model given in equation~\ref{linearinsd}.

\section{Results}

\begin{figure}[!htb]
  \begin{center}
   \includegraphics[scale=1.2,angle=0]{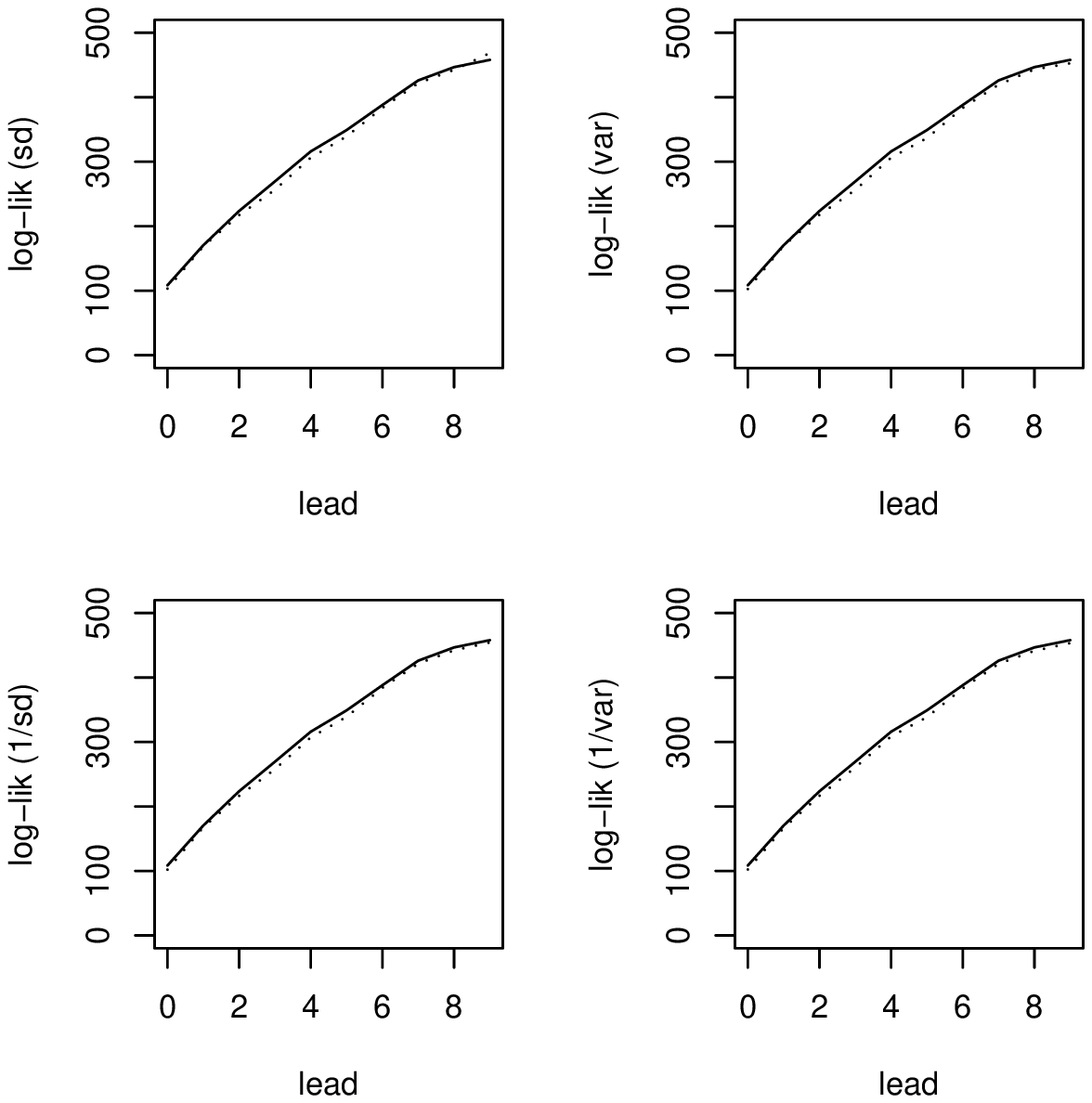}
  \end{center}
  \caption{
The negative log-likelihood scores achieved by a linear regression
(solid line) and four spread-regression models (dotted lines).
  }
  \label{f:f1}
\end{figure}

The first and most important test is to see which of the models achieves
the greatest log-likelihood at the maximum. The results from this test are
shown in figure~\ref{f:f1} (actually in terms of negative log-likelihood
so that smaller is better).
In each case the spread-regression results (dashed lines) are
shown relative to results for a constant-variance model (solid line).
What we see is that the four models achieve roughly the same decrease in
the negative log-likelihood and that in none of the cases is the decrease very large
compared with the change in the log-likelihood from one lead time
to the next.
These changes are also small compared with the
change in the log-likelihood that was
achieved by making the bias correction vary seasonally~\citep{jewson04c}.

\begin{figure}[!htb]
  \begin{center}
   \includegraphics[scale=1.2,angle=0]{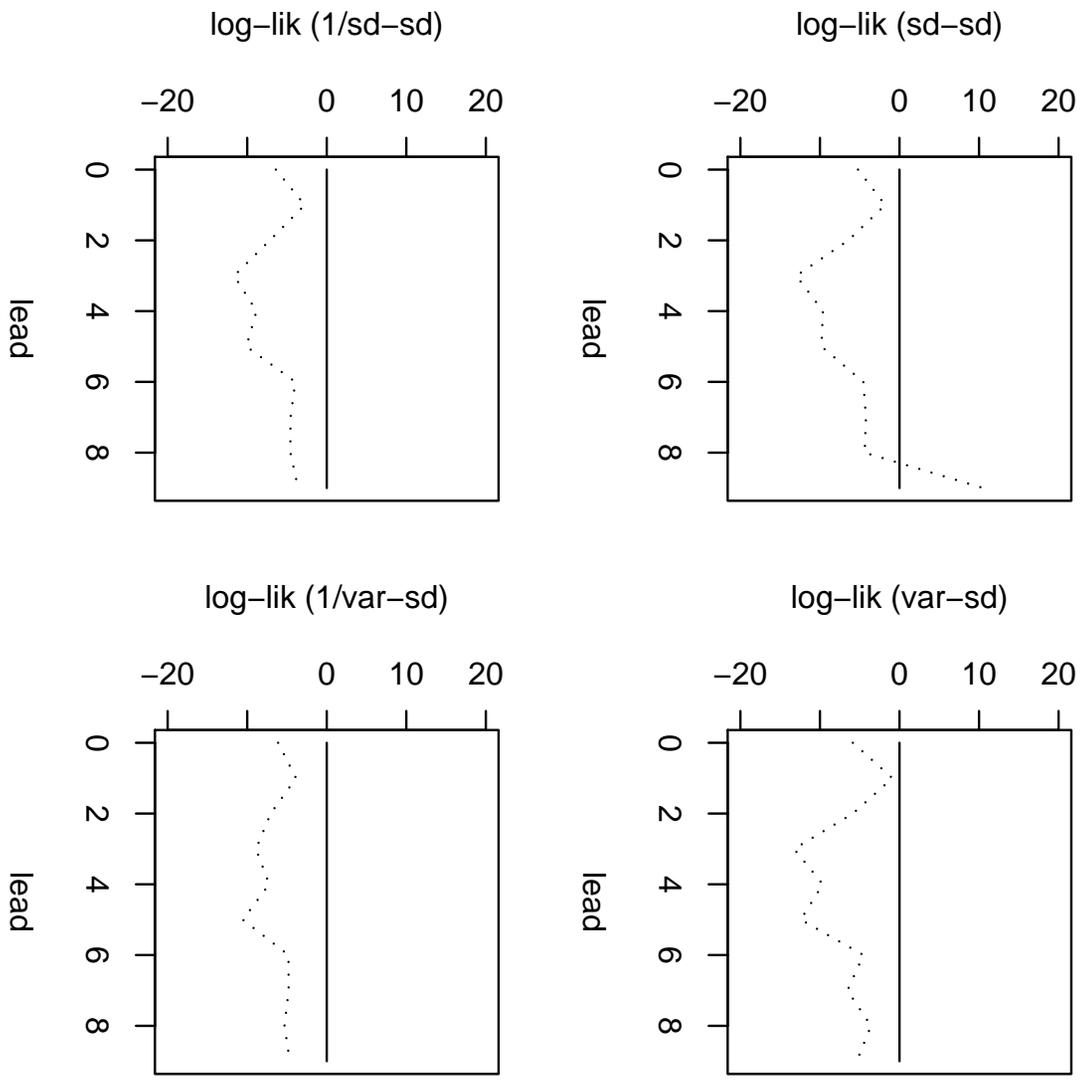}
  \end{center}
  \caption{
As for figure~\ref{f:f1}, but showing the \emph{differences} between
all models and linear regression on a much finer vertical scale.
  }
  \label{f:f2}
\end{figure}

Figure~\ref{f:f2} shows the same data as is shown in figure~\ref{f:f1} but
as differences between the spread-regression models and the constant-variance
model. Again we see that there is little to choose between the models.

\begin{figure}[!htb]
  \begin{center}
   \includegraphics[scale=1.2,angle=0]{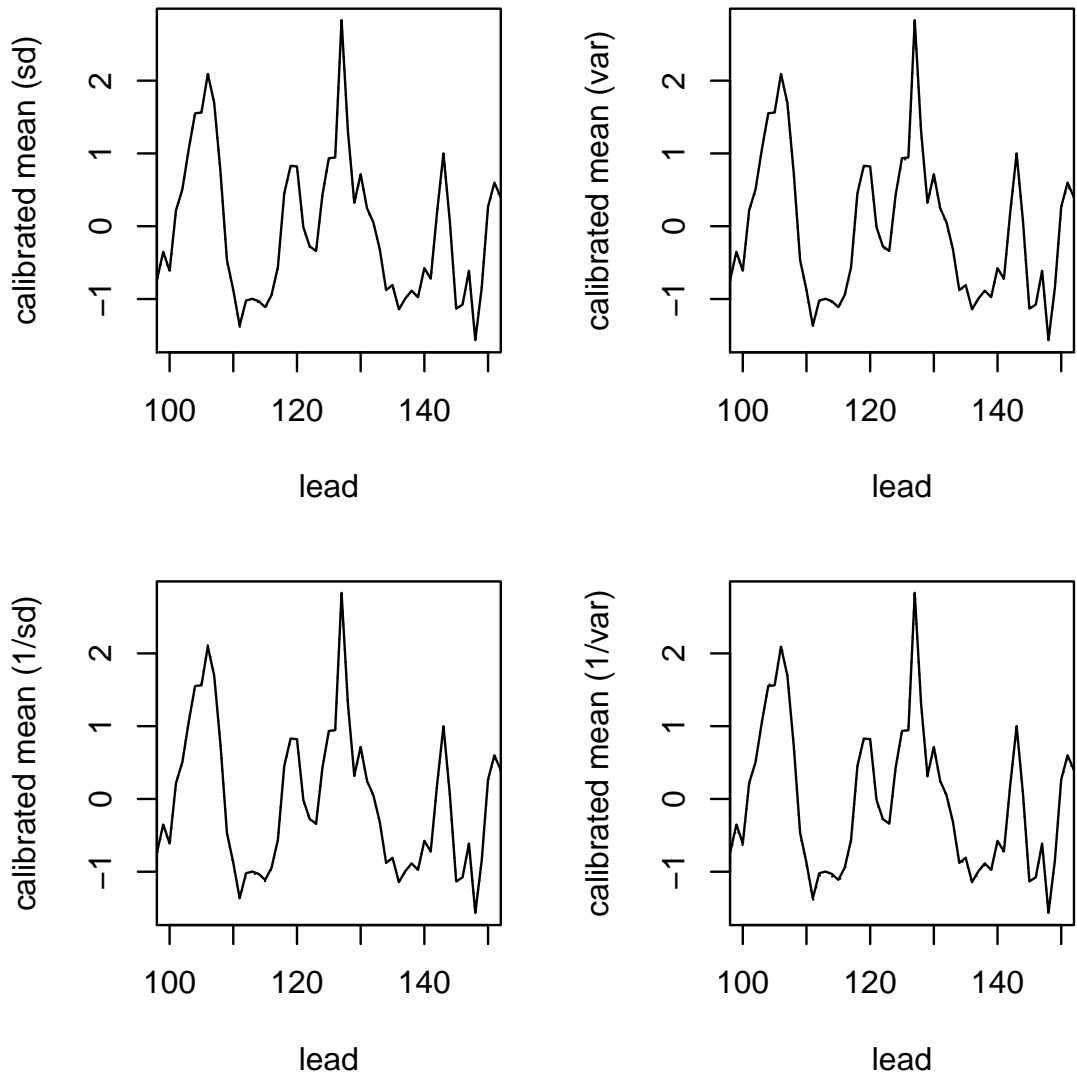}
  \end{center}
  \caption{
  The calibrated mean temperature from linear regression (solid line)
  and four spread-regression models (dotted lines). The dotted lines
  cannot be distinguished because they are so close to the solid lines.
  }
  \label{f:f3}
\end{figure}

\begin{figure}[!htb]
  \begin{center}
   \includegraphics[scale=1.2,angle=0]{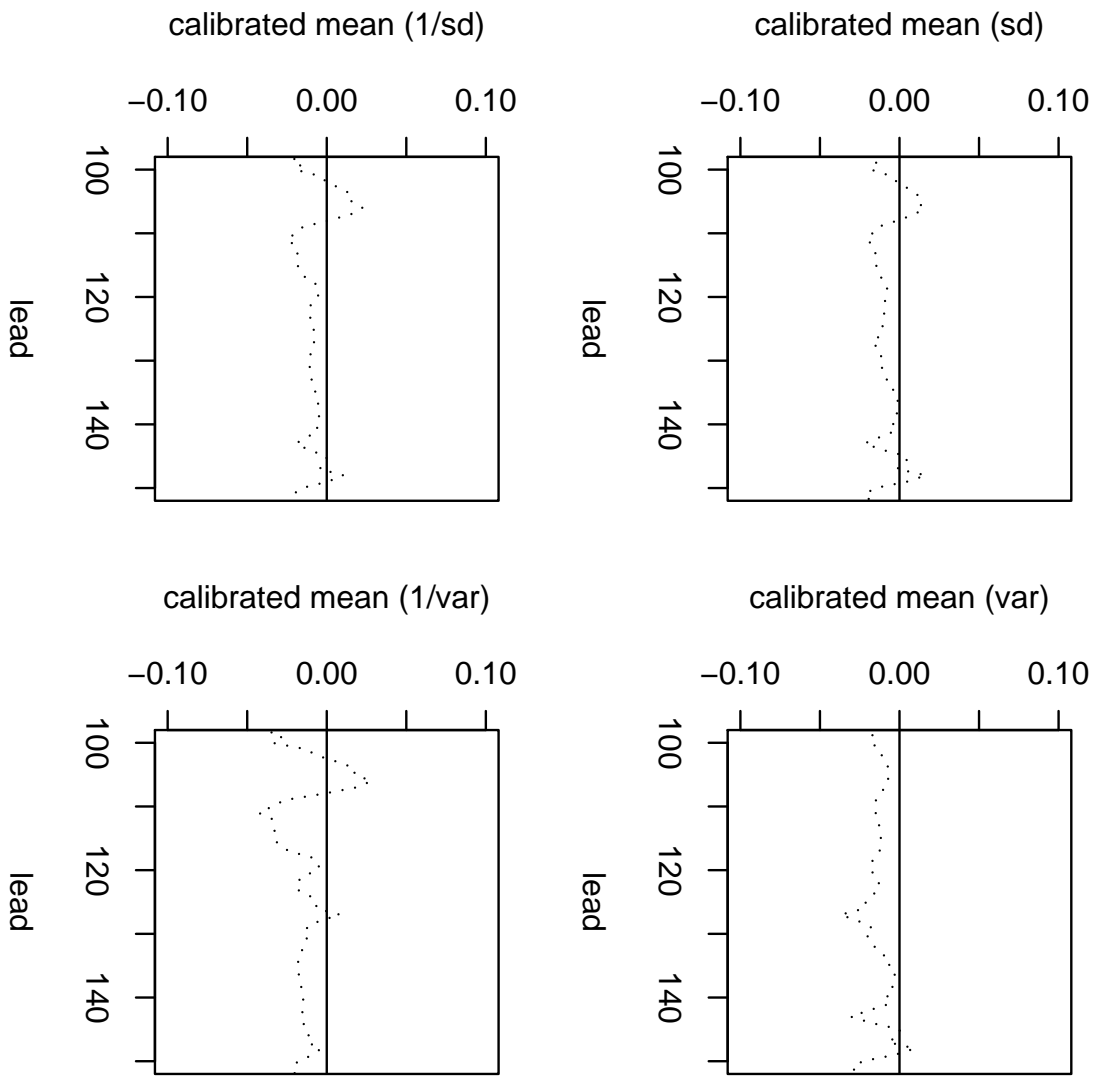}
  \end{center}
  \caption{
As for figure~\ref{f:f4} but showing the \emph{differences} between
all models and linear regression.
  }
  \label{f:f4}
\end{figure}

Figure~\ref{f:f3} shows a fifty-day sample of the calibrated mean
temperature from the constant-variance model with the
spread-regression calibrated temperatures overlaid. The
differences are very small indeed and can only really be seen when
they are plotted explicitly in figure~\ref{f:f4}.

\begin{figure}[!htb]
  \begin{center}
   \includegraphics[scale=1.2,angle=0]{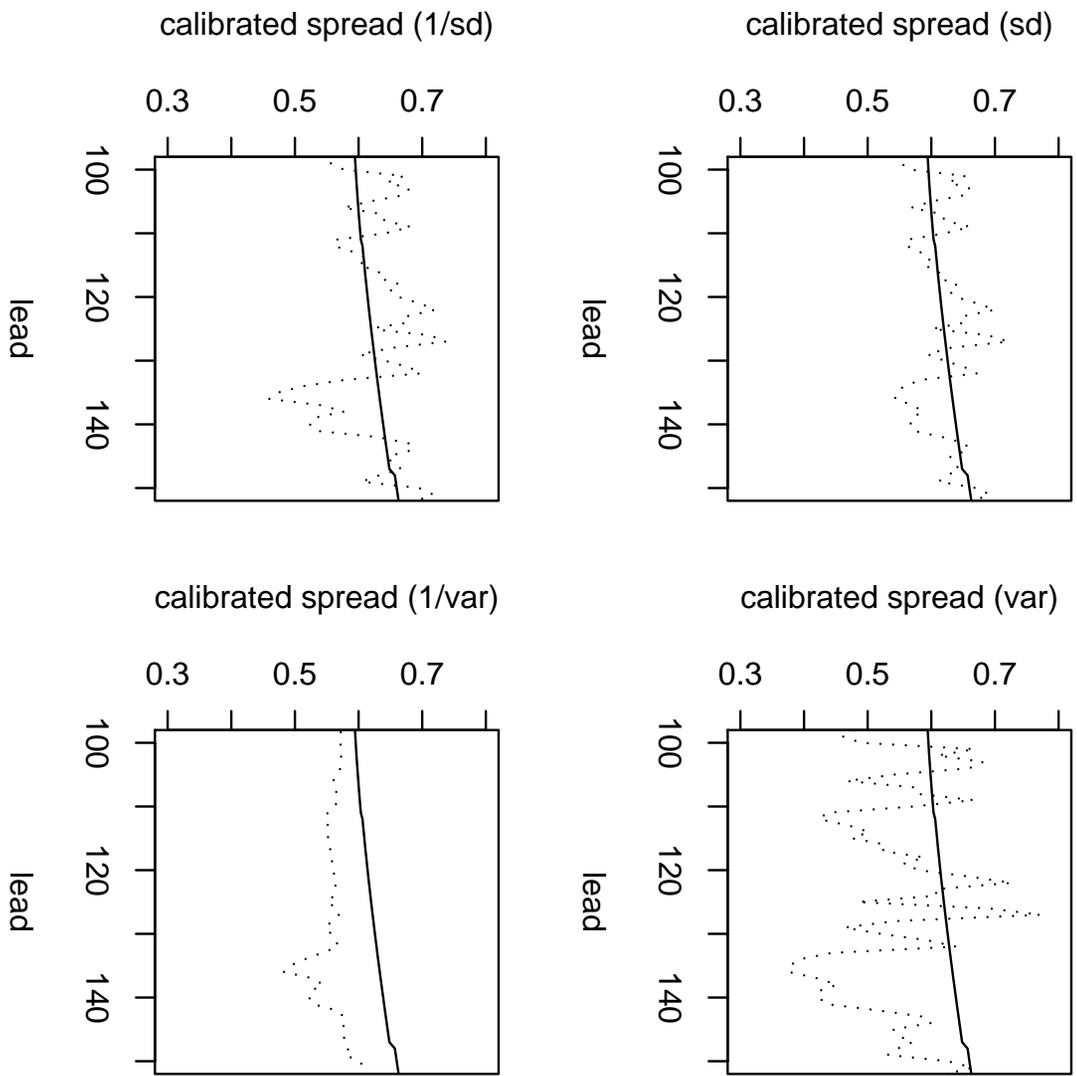}
  \end{center}
  \caption{
  The calibrated uncertainty from linear regression (solid line)
  and four spread-regression models (dotted lines).
  }
  \label{f:f5}
\end{figure}

Figure~\ref{f:f5} shows the calibrated spread from the
constant-variance model and the calibrated spread from the four
spread-regression models.
The uncertainty prediction from the constant variance model varies
slowly from one season to the next and has a kink because of the
presence of missing values in the forecast data.
We now see rather significant
differences between the four spread regression models.
The size of these differences suggests that the variations in $s$ are \emph{not}
so small that the four spread regression models are equivalent to the linear-in-standard-deviation model.

\section{Conclusions}

How to produce good probabilistic temperature forecasts from ensemble
forecasts remains a contentious issue.
This is mainly because of disagreement about how to use the information in the
ensemble spread. We have compared 4 simple parametric models that convert
the spread into an estimate for the forecast uncertainty. All the models allow
for an offset and a term that scales the amplitude of the variability of the
uncertainty. Although the four models lead to visible differences in the
calibrated spread we have found only tiny differences between the impact
of these four models on the log-likelihood achieved.
Also none of the models clearly dominates the others.

These results lead us to conclude that:

\begin{itemize}

    \item the variations in $s$ are not so small that the
calibration of the spread can be linearised, which would make
all four models equivalent

    \item but the changes in the calibrated uncertainty
    \emph{are} small enough that they do not have a great impact on the
    maximum likelihood achieved in any of the models

    \item implying that there is simply not very much information in the variations in the spread

\end{itemize}

It is possible that the models are overfitted to a certain extent. This is unavoidable
given that we only have one year of data for fitting these multiparameter models.
That none of the models dominates is rather curious:
perhaps all the models are equally bad and none of them come close
to modelling the relationship between spread and skill in a reasonable way.
This raises the possibility that better results could perhaps be achieved
by using other parametrisations.

It is difficult to see how to make further progress on these questions until longer series
of stationary back-test data is made available by the numerical modellers.
Meanwhile it seems that a pragmatic approach to producing probabilistic forecasts
would be to stick with the constant variance model since more complex models
have shown only a small benefit in in-sample testing, and do not show a significant
benefit in out-of-sample testing.

\section{Legal statement}

SJ was employed by RMS at the time that this article was written.

However, neither the research behind this article nor the writing
of this article were in the course of his employment, (where 'in
the course of their employment' is within the meaning of the
Copyright, Designs and Patents Act 1988, Section 11), nor were
they in the course of his normal duties, or in the course of
duties falling outside his normal duties but specifically assigned
to him (where 'in the course of his normal duties' and 'in the
course of duties falling outside his normal duties' are within the
meanings of the Patents Act 1977, Section 39). Furthermore the
article does not contain any proprietary information or trade
secrets of RMS. As a result, the authors are the owner of all the
intellectual property rights (including, but not limited to,
copyright, moral rights, design rights and rights to inventions)
associated with and arising from this article. The authors reserve
all these rights. No-one may reproduce, store or transmit, in any
form or by any means, any part of this article without the
authors' prior written permission. The moral rights of the authors
have been asserted.

The contents of this article reflect the authors' personal
opinions at the point in time at which this article was submitted
for publication. However, by the very nature of ongoing research,
they do not necessarily reflect the authors' current opinions. In
addition, they do not necessarily reflect the opinions of the
authors' employers.

\bibliography{jewson}

\clearpage

\clearpage

\clearpage

\clearpage

\clearpage

\end{document}